\documentclass[twocolumn,aps,prl,showpacs,raggedbottom,nobalancelastpage,amssymb,superscriptaddress]{revtex4-1}

\usepackage{graphicx}
\usepackage{grffile}
\usepackage{amsmath}
\usepackage{amssymb}
\usepackage{bm}
\usepackage[usenames]{color}
\usepackage{amsfonts}
\usepackage{appendix}
\usepackage{dsfont}

\def \dH{d^{\text{H}}_{n}}

\def \dF{d^{\text{F}}_{n}}

\def \dS{d^{\text{S}}_{n}}

\begin{document}
\title{Topological Pumping over a Photonic Fibonacci Quasicrystal}
\author{Mor Verbin}
\affiliation{Faculty of Physics, Weizmann Institute of Science, Rehovot 76100, Israel}
\author{Oded Zilberberg}
\affiliation{Institute for Theoretical Physics, ETH Zurich, 8093 Z{\"u}rich, Switzerland}
\author{Yoav Lahini}
\affiliation{Department of Physics, Massachusetts Institute of Technology, Cambridge, MA 02139, USA}
\author{Yaacov E.~Kraus}
\affiliation{Faculty of Physics, Weizmann Institute of Science, Rehovot 76100, Israel}
\author{Yaron Silberberg}
\affiliation{Faculty of Physics, Weizmann Institute of Science, Rehovot 76100, Israel}

\begin{abstract}
Quasiperiodic lattices have recently been shown to be a non-trivial topological phase of matter. Charge pumping -- one of the hallmarks of topological states of matter -- was recently realized for photons in a one-dimensional (1D) off-diagonal Harper model implemented in a photonic waveguide array. The topologically nontrivial 1D Fibonacci quasicrystal (QC) is expected to facilitate a similar phenomenon, but its discrete nature and lack of pumping parameter hinder the experimental study of such topological effects.
In this work we overcome these obstacles by utilizing a family of topologically equivalent QCs which ranges from the Fibonacci QC to the Harper model. Implemented in photonic waveguide arrays, we observe the topological properties of this family, and perform a topological pumping of photons across a Fibonacci QC.
\end{abstract}


\maketitle
%%%%%%%%%%%%%%%%%%%%%%%%%%%%%%%%%%%%%%%%%%%%%%%%%%%%%%%%%%%%%%%%%%

The discovery of topological phases of matter gave birth to an exciting new field of research \cite{First Topological Theory,First Topological Experiment}. The topological classification of gapped systems such as band insulators and superconductors provides insight into the physical behavior of these systems and predicts novel subgap phenomena \cite{Topological Phases}.
One such phenomenon is the topological pump. Making use of the edge states which are the hallmark of any topological system, a dissipationless current of particles across the sample can be generated through an adiabatic change of the parameters of the system  \cite{Original Recipe,Laughlin,Fu,Meidan}. For example, the topologically nontrivial two-dimensional quantum Hall effect (2D QHE) has topologically protected edge states that traverse its energy gaps. Placing it on a cylinder and threading it with an Aharonov-Bohm (AB) flux produces a one-dimensional (1D) quantized charge pump. As the AB flux is continuously increased, an integer number of electrons is transferred across the cylinder for each flux quantum \cite{TKNN,Thouless,Laughlin}.

The 2D QHE is deeply related to the 1D Harper model and its off-diagonal variant  \cite{Harper,AA,Harper Physical Properties}. Whenever the cosine modulation of the model is incommensurate with the underlying lattice, the Harper model describes a quasicrystal (QC), i.e. a lattice which is ordered but non-periodic \cite{Shechtman}. In this case, the AB flux of the 2D QHE becomes equivalent to translations of the 1D Harper model, and boundary states appear in the Harper model as projections of the topologically protected edge states of the 2D QHE. This equivalence suggests that upon a scan of the translational degree of freedom, particles would be pumped from one boundary to the other.

In a recent experiment, the off-diagonal Harper model was implemented in a photonic waveguide array \cite{Original Recipe}. The quasiperiodic cosine modulation of the model was produced by controlling inter-waveguide distances. By adiabatically varying the relative phase between the modulation and the underlying lattice, light was pumped across the sample, revealing its topological nature. Producing a deep connection between topological phases of matter and the seemingly unrelated topic of QCs, this work generated growing interest in the boundary phenomena of QCs \cite{Citing 1, Citing 2, Citing 3, Citing 4, Citing 5}. One development was the discovery of the topological origin of the localized boundary modes of the Fibonacci QC \cite{Smoothening} -- a binary QC whose lattice spacings are two discrete values that appear interchangeably according to the Fibonacci sequence.
Despite the topological properties of this QC, when studying its subgap boundary modes it becomes apparent that an adiabatic pumping will be challenging. As the model does not include an obvious pumping parameter, and adiabatic processes cannot be done in its discrete potential, it is unclear whether this theoretically-proposed process can be realized experimentally.

In this Letter, we report a topological pump over a Fibonacci QC implemented in a photonic waveguide array. To achieve this, we harness the recently found topological equivalence between the Fibonacci QC and the Harper model. This equivalence can be accessed through a single deformation parameter, whose range spans an extensive family of topologically equivalent QC models \cite{Smoothening}. We thereby perform a two-parameter topological pumping that includes (i) a deformation of the Fibonacci QC to a smoothened topologically equivalent model, (ii) a scan of the translation parameter, and (iii) a deformation back into a Fibonacci QC. Thus, this Letter contains an experimental demonstration of several fundamental concepts, including the deep relationship between topological pumps and QCs, the topological equivalence between different QCs, and the connection between their boundary phenomena.

Photonic waveguide arrays serve as a highly versatile and customizable platform to study the properties of topological pumps and of QCs \cite{Original Recipe,Phase transitions}. 
In these arrays, evanescent coupling between adjacent waveguides allows photons to hop from one waveguide to the other along the propagation axis, denoted by $z$. The resulting dynamics of light propagation is described by a Schr\"{o}dinger equation with $z$ taking over the role of time, $i\partial_{z}\psi_{n}=H\psi_{n}$, where $\psi_{n}$ is the wavefunction at waveguide number $n$. $H$ is a general off-diagonal tight-binding Hamiltonian
\begin{align}
H\psi_{n} & =t_{n}\psi_{n-1}+t_{n+1}\psi_{n+1} \, ,\label{Eq:H_TBM}
\end{align}
where $t_{n}$ is the hopping amplitude from waveguide $n$ to waveguide $n-1$.

To introduce quasiperiodicity into the system, the values of $t_{n}$ are modulated according to
\begin{align} \label{Eq:Hopping}
t_n & =t_{0}\left[1+\lambda d_n\right] \, ,
\end{align}
where $t_{0}$ is the characteristic hopping amplitude of the system, $\lambda \in [0,1)$ is the modulation strength, and $d_n \in [-1,1]$ can be any chosen quasiperiodic modulation function.

In the off-diagonal Harper model, the quasiperiodicity enters in the form of a cosine modulation
\begin{align} \label{Eq:Hopping_Harper}
\dH=\cos(2\pi bn+\phi) \,.
\end{align}
This model's long-range order originates from the cosine function, and is controlled by the modulation frequency $b$ \cite{Han}. A QC is produced whenever the hopping modulation is incommensurate with the underlying lattice (i.e. $b$ is irrational). Correspondingly, the parameter $\phi$ shifts the origin of the modulation. This shift degree of freedom is equivalent to the AB flux in the 2D QHE and spans a family of models that corresponds to a topological pump \cite{Thouless,Original Recipe}. Accordingly, topological boundary states appear and disappear as a function of $\phi$ \cite{Phase transitions}.

Comparably, a Fibonacci-like QC is constructed of a sequence of two distinct values which are ordered in a quasiperiodic manner:
\begin{equation} \label{Eq:d_Fibonacci}
\resizebox{.85\hsize}{!}{$\displaystyle \dF=2\left(\left\lfloor \frac{\tau}{\tau+1}\left(n+2\right)\right\rfloor -\left\lfloor
\frac{\tau}{\tau+1}\left(n+1\right)\right\rfloor \right)-1=\pm1 \, ,$}
\end{equation}
where $\lfloor x \rfloor$ is the floor function. This sequence is obtained by applying the ``cut-and-project'' procedure, i.e. projecting a strip of a square lattice onto the line $y=x/\tau$ \cite{QC Senechal}. Whenever the slope of the line, $\tau$, is irrational, Eq.~\eqref{Eq:d_Fibonacci} becomes quasiperiodic. For example, the case of $\tau = (1+\sqrt{5})/2$ is the well-known Fibonacci QC \cite{Fibonacci Physical Properties}. The absence of a shift parameter $\phi$ in Eq.~\eqref{Eq:d_Fibonacci} is the first obstacle in our attempt to use this model for topological pumping.

The Harper and Fibonacci models have different physical properties \cite{Harper Physical Properties, Fibonacci Physical Properties}, and until recently, only partial success has been achieved in attempts to combine them under the same general framework \cite{Hiramoto and Kohmoto, Unification Theory 1, Unification Theory 2}. A recent paper presented a smooth deformation between the two models, which preserves the topological properties of their energy spectra and enables the definition of a generalized family of topologically equivalent QCs \cite{Smoothening}:
\begin{equation} \label{eq:d_Smooth}
\resizebox{.85\hsize}{!}{$\displaystyle
\dS(\beta)=\frac{\tanh\left\{
\beta\left[\cos\left(2\pi\bar{b}\cdot\frac{2n+3}{2}+\phi\right)-\cos\left(\pi\bar{b}\right)\right]\right\}
}{\tanh\left(\beta\right)} \, .$}
\end{equation}
At the limit of $\beta\rightarrow 0$, $\tanh\left(\beta x\right)/\tanh\left(\beta\right)\rightarrow x$, yielding the Harper modulation with $\bar{b}=b$, i.e. $\dS\rightarrow \dH$, up to a constant shift. At the opposite limit of $\beta\rightarrow\infty$, $\tanh\left(\beta x\right)/\tanh\left(\beta\right)\rightarrow\textrm{sign}(x)$, so for $\bar{b}=(\tau+1)/\tau$ we obtain a Fibonacci QC, i.e. $\dS\rightarrow \dF$.

The topological class (i.e. the Chern number) of a gap in the energy spectrum which remains open along the deformation is independent of $\beta$ \cite{Smoothening, Dana}. This means that different QCs with the same irrational modulation frequency $\bar{b}$ are topologically equivalent as long as their gaps remain open as a function of $\beta$. An important outcome of this deformation is that the parameter $\phi$ now gives us a way of incorporating an equivalent of the AB-flux into the Fibonacci QC. It acts as a controllable knob to experimentally observe boundary states and to generate a topological pump \cite{Smoothening}.

\begin{figure}[!ht]
\includegraphics[width=3.38in]{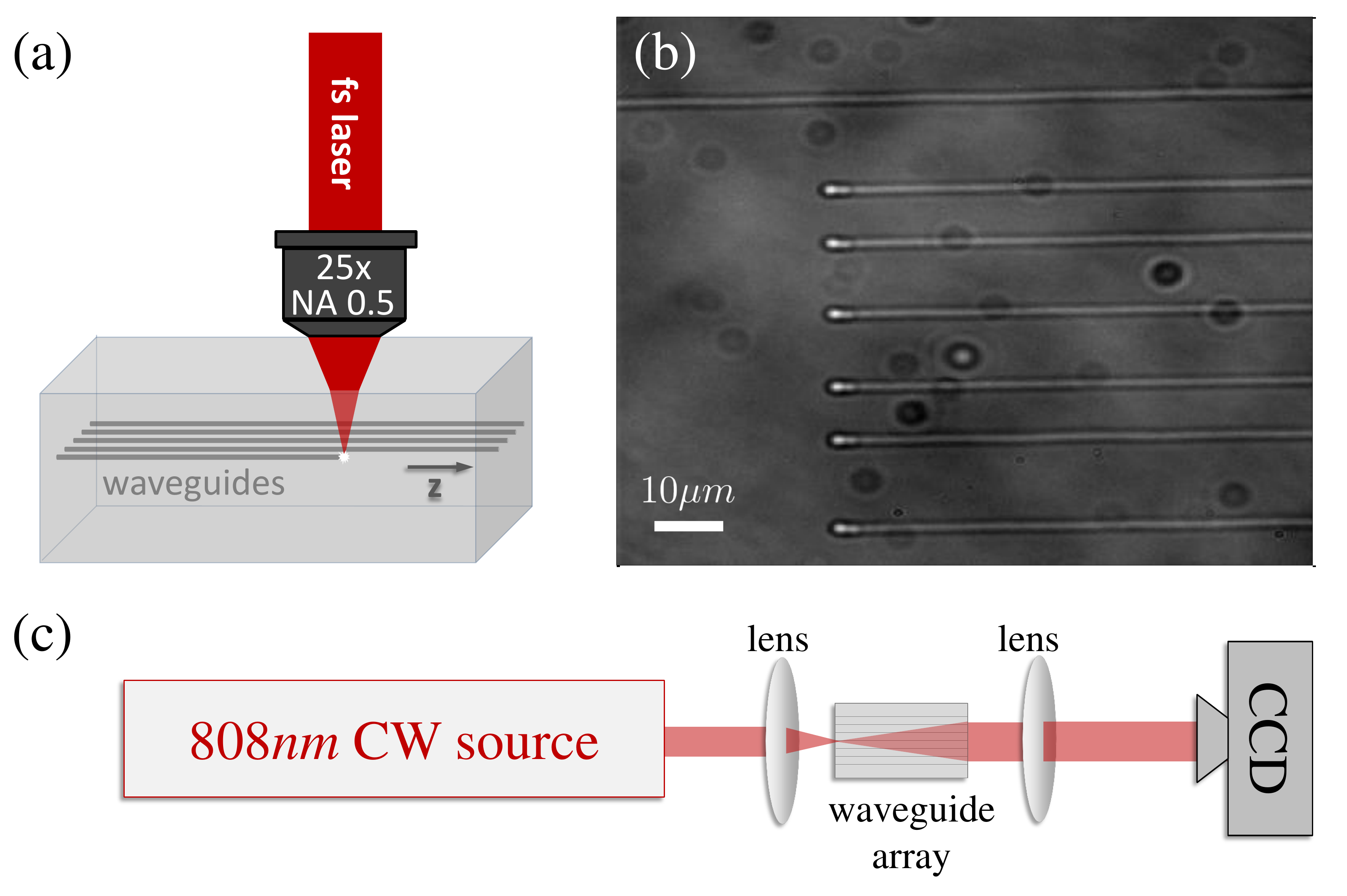}
\caption{\label{Figure1} (color online) Experimental Setup. (a) Pulses from a femtosecond fiber laser (Raydiance Smart Light) with a central wavelength of $1552 nm$ are focused inside the glass, resulting in a permanent change to the index of refraction. 
(b) Microscope image of a waveguide array used in one of the experiments. One waveguide runs from one side of the glass slide to the other and is used for input. The rest of the waveguides start at some distance from input facet of the glass, with the distance determining the propagation length along the array. Inset marker is $10 \mu m$ long, while the distances between the waveguides range from $8 \mu m$ to $12 \mu m$. (c) A $808nm$-wavelength beam from a continuous-wave (CW) diode laser is injected into the waveguide array, allowed to propagate along it, and measured at the output using a CCD camera.}
\end{figure}

To study the topological properties of this generalized family of QCs, waveguide arrays are fabricated in 75mm-long bulk glass slides using femtosecond-laser microfabrication technique, as illustrated in Fig.~\ref{Figure1}(a) \cite{Szameit}. The waveguides are identical in both refractive index and size, and the separation between them is customized for the desired quasiperiodic profile of coupling coefficients. The resulting structure is visible to an optical microscope, as seen in Fig.~\ref{Figure1}(b). To study the dynamics of light propagating within the system, a continuous-wave laser beam is injected into one of the waveguides in the array, allowed to propagate along it, and measured at the output, giving the end-result of the dynamics, as illustrated in Fig.~\ref{Figure1}(c).

%%%%%%%%%%%%%%%%%%%%%%%%%%%%%%%%%%%%%%%%%%%%%%%%%%%%%%%%%%%%%%%%
%%%%%%%%%%%%%%%%%%%%%%%%%%%%%%%%%%%%%%%%%%%%%%%%%%%%%%%%%%%%%%%%

\begin{figure}[!ht]
\includegraphics[clip]{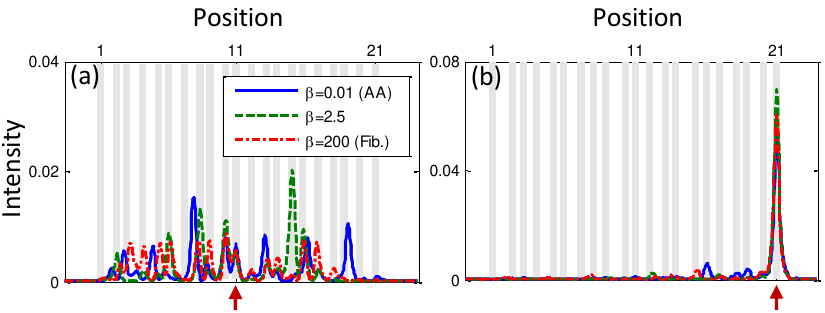}
\caption{\label{Figure2} (color online) Experimental observation of the right boundary state for 21-waveguide-long arrays with $\beta=0.01$ (a Harper QC), $\beta=2.5$ and $\beta=200$ (a Fibonacci QC), with $\lambda=0.6$, $\bar{b}=(1+\sqrt{5})/2$ and $\phi=0.7\pi$. Light was initially injected into a single waveguide (red arrows). The measured outgoing intensity is plotted versus the waveguide position. (a) An excitation at the middle of the array (site 11) results in a significant spread of the wavefunction. (b) Regardless of the value of $\beta$, for an excitation at the rightmost waveguide (site 21) the light remains tightly localized at the boundary, marking the existence of a boundary state.}
\end{figure}

While the bulk properties of the Harper and Fibonacci models have been studied extensively \cite{Fibonacci Physical Properties,Harper Physical Properties}, their boundary states have received less attention. 
Our first experiment is designed to observe boundary states in the generalized QC model presented in Eq.~\eqref{eq:d_Smooth}. For this purpose three arrays were fabricated with $\beta=0.01$ (a Harper QC), $\beta=2.5$, and $\beta=200$ (a Fibonacci QC). The resulting experimental observations are depicted in Fig.~\ref{Figure2}. In all three arrays, light injected into a waveguide in the middle of the array showed significant expansion due to the overlap of the input light with the extended bulk eigenstates of the system. However, when light was injected into the rightmost waveguide, the intensity distribution remained tightly localized at the boundary, revealing the existence of a localized boundary state in all three arrays.
This result accentuates the connection between the boundary states of the Fibonacci QC and those of the Harper model, showing their topological origin \cite{A, B, C, D}.

%%%%%%%%%%%%%%%%%%%%%%%%%%%%%%%%%%%%%%%%%%%%%%%%%%%%%%%%%%%%%%%%
%%%%%%%%%%%%%%%%%%%%%%%%%%%%%%%%%%%%%%%%%%%%%%%%%%%%%%%%%%%%%%%%

\begin{figure}[!ht]
\includegraphics[clip]{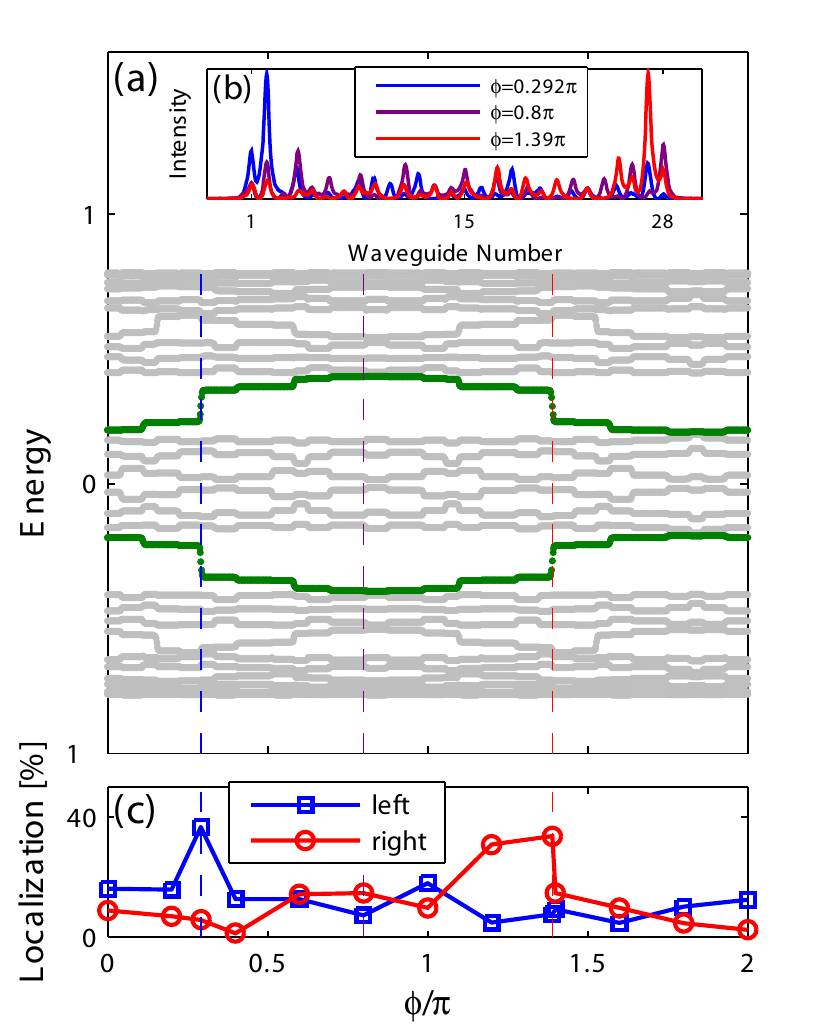}
\caption{\label{Figure3} (color online) Spectrum and wavefunction localization vs. $\phi$. (a) Energy spectrum of a 28-long Fibonacci QC ($\beta=100$) vs. $\phi$, with $\lambda=0.25$ and $\bar{b}=(1+\sqrt{5})/2$. The two boundary states that traverse the largest energy gaps as a function of $\phi$ are marked in green. Blue, purple and red mark three values of $\phi$ ($0.292\pi$, $0.8\pi$ and $1.39\pi$ correspondingly) for which experimental results are presented in the inset. (b) Output intensity for light inserted into the (blue) leftmost waveguide, exciting the left-hand-side boundary state; (purple) middle waveguide, when the state is found within the energy band; (red) rightmost waveguide, exciting the right-hand-side boundary state. (c) The amount of light remaining at the two outer-most waveguides vs. $\phi$ on the (blue) left-hand and (red) right-hand boundaries.}
\end{figure}

In a second set of experiments, we studied the effect of the parameter $\phi$. As mentioned above, in the Harper QC ($\beta\rightarrow 0$), $\phi$ shifts the location of the boundary state from the right to the left boundary \cite{Original Recipe}. To study the other limit, at $\beta=100$, thirteen waveguide arrays have been fabricated, for different values of $\phi$ between $0$ and $2\pi$. Figure \ref{Figure3}(a) depicts the numerically obtained energy spectrum of the Hamiltonian of this system as a function of $\phi$. The spectrum is broken into a set of bands and gaps which remains mostly unchanged, but includes two states that counter-traverse the largest energy gaps as a function of $\phi$. When found within the gaps, these states are localized at either the left or right boundary of the system. Light inserted into the corresponding boundaries excites these boundary states and remains there. When these eigenstates are located within the energy band, they behave as bulk states. Accordingly, inserted light spreads across the array, as seen in Fig. \ref{Figure3}(b). The amount of light which remained at the two outer-most waveguides (closest to the injection sites) as a function of $\phi$ is presented in Fig.~\ref{Figure3}(c). The two observed peaks correspond to the values of $\phi$ for which boundary states exist on the same side of the array where light was injected. These results show that the dependence of the Fibonacci QC on the parameter $\phi$ is similar to that of the Harper model. It should therefore allow us to perform a topological pumping of photons from one side of a Fibonacci QC to the other.

%%%%%%%%%%%%%%%%%%%%%%%%%%%%%%%%%%%%%%%%%%%%%%%%%%%%%%%%%%%%%%%%
%%%%%%%%%%%%%%%%%%%%%%%%%%%%%%%%%%%%%%%%%%%%%%%%%%%%%%%%%%%%%%%%

\begin{figure}[!ht]
\includegraphics[clip]{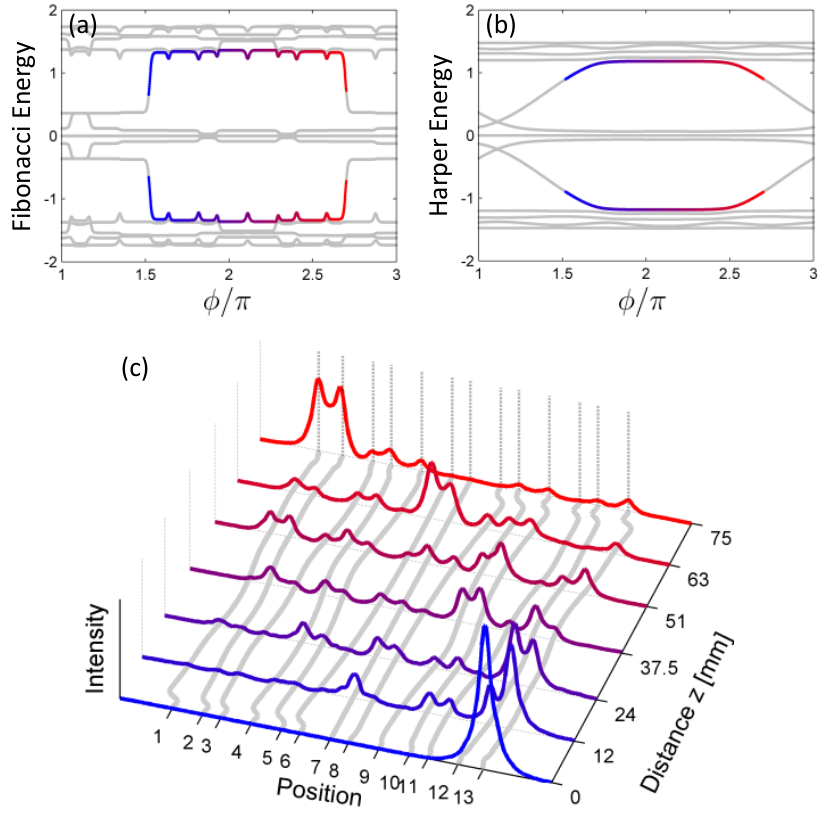}
\caption{\label{Figure4} (color online) Two-parameters topological pumping. (a-b) Energy spectra as a function of $\phi$, of a 13-waveguide-long (a) Fibonacci QC ($\beta=200$) and (b) Harper QC ($\beta=0.01$) with $\lambda=0.6$. (c) Experimental results of topological pumping. Light was injected into the rightmost waveguide (site 13) of a Fibonacci QC ($\beta=200$) at $z=0$. An adiabatic process lowers $\beta$ to $0.01$, resulting in a Harper QC at $z=5 mm$. $\phi$ is then scanned from $1.52\pi$ to $2.7\pi$ to pump light across the structure. Finally, $\beta$ is increased back to $200$ between $z=70 mm$ and $z=75 mm$, returning to the Fibonacci QC. The measured intensity distributions as a function of the position are presented at different stages of the adiabatic evolution, i.e., different propagation distances along the slide. The light is pumped across the array from right to left, ending up localized at the two leftmost waveguides (sites 1 and 2).}
\end{figure}

% For my last trick...
Ideally, pumping could be done in the Fibonacci QC by adiabatically scanning $\phi$ to allow light to follow the localized state from one side of the array to the other \cite{Original Recipe}. However, a comparison of the energy spectra of the Fibonacci and Harper QCs (see Figs.~\ref{Figure4}(a) and (b), respectively), reveals an experimental obstacle. As $\beta$ increases, the region in which the boundary states traverse the gaps becomes shorter, and as we approach the Fibonacci QC they become infinitesimally small \cite{Immediate transition}. This sharp traversal hinders any adiabatic processes, as these requires an infinite propagation length to adiabatically follow the localized state from the energy gap to the energy band. Notwithstanding, this problem can be circumvented by the topological equivalence maintained along the deformation in Eq.~\eqref{eq:d_Smooth}. Since the topological class of the largest energy gaps of the QC is independent of $\beta$, we can start with an adiabatic deformation of the Fibonacci QC into a Harper QC by decreasing $\beta$, then adiabatically scan $\phi$ for the pumping process, and end with another adiabatic deformation back into a Fibonacci QC.

%%%%%%%%%%%%%%%%%%%%%%%%%%%%%%%%%%%%%%%%%%%%%%%%%%%%%%%%%%%%%%%%
%%%%%%%%%%%%%%%%%%%%%%%%%%%%%%%%%%%%%%%%%%%%%%%%%%%%%%%%%%%%%%%%

This procedure was implement using 7 waveguide arrays with the same scans of $\beta$ and $\phi$, but of increasing propagation length. This allows allowing the observation of different steps along the adiabatic propagation. The experimental results are shown in Fig.~\ref{Figure4}(c). The reconstruction of the propagation shows the initially excited localized state leaving the right boundary of the Fibonacci QC, expanding along the waveguide array as it propagates, and re-localizing at the other side. This flow of light across the array is a genuine topological pumping of photons.

To conclude, in this Letter we have experimentally studied and verified the topological properties of a generalized family of QC which ranges from the Harper model to the Fibonacci QC. All the members of this family have boundary states that appear and disappear as a function of the translation degree-of-freedom, $\phi$. Members of this family which can be deformed into each other while keeping the main energy gaps open are topologically equivalent. Using this equivalence, we circumvent the obstacles which arise when trying to perform a topological pumping across a Fibonacci QC, making it the first time the mathematical notion of topological equivalence is utilized to solve an experimental problem. This paper further develops the ongoing research of topological phases of matter using quasiperiodic photonic arrays.

We thank the Minerva Foundation, U.S. Army RFEC-Atlantic, the Swiss National Science Foundation, the Pappalrado Fellowship in Physics, and the ICore program of the ISF for financial support.


\begin{thebibliography}{32}
\expandafter\ifx\csname natexlab\endcsname\relax\def\natexlab#1{#1}\fi \expandafter\ifx\csname bibnamefont\endcsname\relax
  \def\bibnamefont#1{#1}\fi
\expandafter\ifx\csname bibfnamefont\endcsname\relax
  \def\bibfnamefont#1{#1}\fi
\expandafter\ifx\csname citenamefont\endcsname\relax
  \def\citenamefont#1{#1}\fi
\expandafter\ifx\csname url\endcsname\relax
  \def\url#1{\texttt{#1}}\fi
\expandafter\ifx\csname urlprefix\endcsname\relax\def\urlprefix{URL }\fi \providecommand{\bibinfo}[2]{#2}
\providecommand{\eprint}[2][]{\url{#2}}

\bibitem{First Topological Theory}
C. L. Kane and E. J. Mele, Phys. Rev. Lett. 95, 146802 (2005).

\bibitem{First Topological Experiment}
M. König et al., Science 318, 766–770 (2007).

\bibitem{Topological Phases}
M. Z. Hasan and C. L. Kane, Rev. Mod. Phys. 82, 2045 (2010); X.-L. Qi and S.-C. Zhang, ibid. 83, 1057 (2011).

\bibitem{Original Recipe}
K.E. Kraus, Y. Lahini, Z. Ringel, M. Verbin, and O. Zilberberg, Phys. Rev. Lett. 109, 106402 (2012).

\bibitem{Fu}
L.~Fu and C.~L. Kane, Phys. Rev. B 74, 195312 (2006).

\bibitem{Meidan}
D.~Meidan, T.~Micklitz, and P.~W. Brouwer, Phys. Rev. B 82, 161303 (2010); \textit{ibid.} 84, 195410 (2011).

\bibitem{Laughlin}
R. B. Laughlin, Phys. Rev. B 23, 5632 (1981).

\bibitem{Thouless}
D.J. Thouless, Phys. Rev. B 27, 6083 (1983).

\bibitem{TKNN}
D.~J. Thouless, M. Kohmoto, M.~P. Nightingale, and M. den Nijs, Phys. Rev. Lett. 49, 405 (1982).

\bibitem{Harper}
P.G. Harper, Proc. Phys. Soc. London A 68, 874 (1955).

\bibitem{AA}
S. Aubry and G. Andr\'{e}, Ann. Isr. Phys. Soc. 3, 133 (1980).

\bibitem{Harper Physical Properties}
J. H. Han, D. J. Thouless, H. Hiramoto, and M. Kohmoto, Phys. Rev. B 50, 11365 (1994).

\bibitem{Shechtman}
D. Shechtman, I. Blech, D. Gratias, J. W. Cahn, Phys. Rev. Lett. 1984, 53, 1951–1953.

\bibitem{Citing 1}
L.J. Lang, X. Cai, and S. Chen, Phys. Rev. Lett. 108, 220401 (2012).

\bibitem{Citing 2}
F. Mei, S.L. Zhu, Z.M. Zhang, C.H. Oh, and N. Goldman, Phys. Rev. A 85, 013638 (2012).

\bibitem{Citing 3}
Sriram Ganeshan, Kai Sun and S. Das Sarma, Phys. Rev. Lett 110, 180403 (2013).

\bibitem{Citing 4}
Indubala I. Satija and Gerardo G. Naumis, Phys. Rev. B 88, 054204 (2013).

\bibitem{Citing 5}
O. Viyuela, A. Rivas, and M. A. Martin-Delgado, Phys. Rev. B 86, 155140 (2013).

\bibitem{Smoothening}
K.E. Kraus, and O. Zilberberg, Phys. Rev. Lett. 109, 116404 (2012).

\bibitem{Phase transitions}
M. Verbin, O. Zilberberg, K.E. Kraus, Y. Lahini and Y. Silberberg, Phys. Rev. Lett. 110, 076403 (2013).

\bibitem{Han}
J. H. Han and D. J. Thouless, Phys. Rev. B 50, 11365 (1994).

\bibitem{QC Senechal}
M. Senechal, Quasicrystals and Geometry (Cambridge University Press, Cambridge, England, 1996).

\bibitem{Fibonacci Physical Properties}
M. Kohmoto, L. P. Kadanoff, and C. Tang, Phys. Rev. Lett. 50, 1870 (1983); S. Ostlund, R. Pandit, D. Rand, H. J. Schellnhuber, and E. D. Siggia, Phys. Rev. Lett. 50, 1873 (1983).

\bibitem{Hiramoto and Kohmoto}
H. Hiramoto and M. Kohmoto, Int. J. Mod. Phys. B 06, 281 (1992).

\bibitem{Unification Theory 1}
H. Hiramoto and M. Kohmoto, Phys. Rev. Lett. 62, 2714 (1989).

\bibitem{Unification Theory 2}
G. G. Naumis, F. J. López-Rodríguez, Physica B 403, 1755-1762 (2008).

\bibitem{Dana}
I. Dana, arXiv:1310.7970.

\bibitem{Szameit}
A. Szameit, D. Blomer, J. Burghoff, T. Schreiber, T. Pertsch, S. Nolte, A. Tunnermann, and F. Lederer, Opt. Express 13, 10552 (2005).

\bibitem{A}
E. S. Zijlstra, A. Fasolino, and T. Janssen, PRB 59, 302 (1999).

\bibitem{B}
Y. El Hassouani et al., PRB 74, 035314 (2006).

\bibitem{C}
X.-N. Pang, J.-W. Dong, and H.-Z. Wang, J. Opt. Soc. Am. B 27, 2009 (2010).

\bibitem{D}
A. J. Martínez and M. I. Molina, PRA 85, 013807 (2012).

\bibitem{Immediate transition}
This transition is immediate for a Fibonacci QC at $\beta\rightarrow\infty$, where no gap-traversing topological boundary state exists.


\end{thebibliography}
\end{document}